\documentclass[useAMS,usenatbib]{mn2e}

\usepackage{graphicx}
\usepackage{epsfig}
\usepackage{amssymb}

\title[Long-Term Stability of Horseshoe Orbits]{Long-Term Stability of Horseshoe Orbits}
\author[Matija \' Cuk, Douglas P. Hamilton and Matthew J. Holman]{Matija \' Cuk$^{1}$\thanks{E-mail:mcuk@seti.org (MC)}, Douglas P. Hamilton$^{2}$ and Matthew J. Holman$^{3}$\\
$^{1}$Carl Sagan Center, SETI Institute, Mountain View, CA 94043, USA\\
$^{2}$Department of Astronomy, University of Maryland, College Park, MD 20742, USA\\
$^{3}$Smithsonian Astrophysical Observatory, Cambridge, MA 02138, USA}
\begin{document}

\date{June 7, 2012}

\pagerange{\pageref{firstpage}--\pageref{lastpage}} \pubyear{2012}

\maketitle

\label{firstpage}

\begin{abstract}
Unlike Trojans, horseshoe coorbitals are not generally considered to be long-term stable \citep{der81a, md99}. As the lifetime of Earth's and Venus's horseshoe coorbitals is expected to be about a Gyr, we investigated the possible contribution of late-escaping inner planet coorbitals to the lunar Late Heavy Bombardment. Contrary to analytical estimates, we do not find many horseshoe objects escaping after the first 100 Myr. In order to understand this behaviour, we ran a second set of simulations featuring idealized planets on circular orbits with a range of masses. We find that horseshoe coorbitals are generally long lived (and potentially stable) for systems with primary-to-secondary mass ratios larger than about 1200. This is consistent with the results of \citet{lau02} for equal-mass pairs of coorbital planets and the instability of Jupiter's horseshoe companions \citep{sta08}. Horseshoe orbits at smaller mass ratios are unstable because they must approach within 5 Hill radii of the secondary. In contrast, tadpole orbits are more robust and can remain stable even when approaching within 4 Hill radii of the secondary.
\end{abstract}


\section{Introduction}

A horseshoe orbit is a type of coorbital motion where two bodies maintain the same average distance from the central object, while librating around points $180^{\circ}$ apart in longitude. If one body is much more massive than the other, the smaller object's trajectory resembles a horseshoe when plotted in the frame rotating with the bodies' average mean motion. Horseshoe orbits were a little-known theoretical prediction \citep{bro11, rab61} until it was realized that Saturn's moons Janus and Epimetheus actually follow such paths \citep{smi80}. 

The most detailed theoretical exploration of horseshoe orbits was undertaken by \citet{der81a, der81b}, and the same treatment is featured in the textbook by \citet{md99}. Using an analytical approach, \citet{der81a} could not confirm that successive horseshoe loops repeat exactly, even in the restricted problem. Instead, they derived an estimate of possible drift during one horseshoe cycle, and extrapolated the lifetime of the horseshoe configuration assuming that these drifts over long periods of time behave like a random walk. With this approach, \citet{der81a} derived the following formula for the horseshoe lifetime
\begin{equation}
\tau \lesssim T / \mu^{5/3},
\label{tau}
\end{equation}
where $T$ the orbital period and $\mu$ is the secondary-to-primary mass ratio. While this formula was originally derived as a first-order estimate of how long the horseshoe motion may last (S. Dermott, 2010, personal communication), it was later presented as a relatively firm upper limit on the lifetime of such configurations \citep[e.g.][] {md99, sta08, chr11}. Note that this relation indicates that the observed horseshoe coorbitals Janus and Epimetheus can last longer than 10 Gyr, while Jovian horseshoes should be unstable on Myr timescales \citep[as confirmed numerically by ][]{sta08}. Given that this relation is based on a relatively simple extrapolation, the ultimate intrinsic stability of horseshoe orbits must be explored using direct integrations which were not possible in the Voyager era.

\begin{figure*}
\begin{minipage}{2.8truein}
\centering
(a)\includegraphics[scale=.5, angle=270]{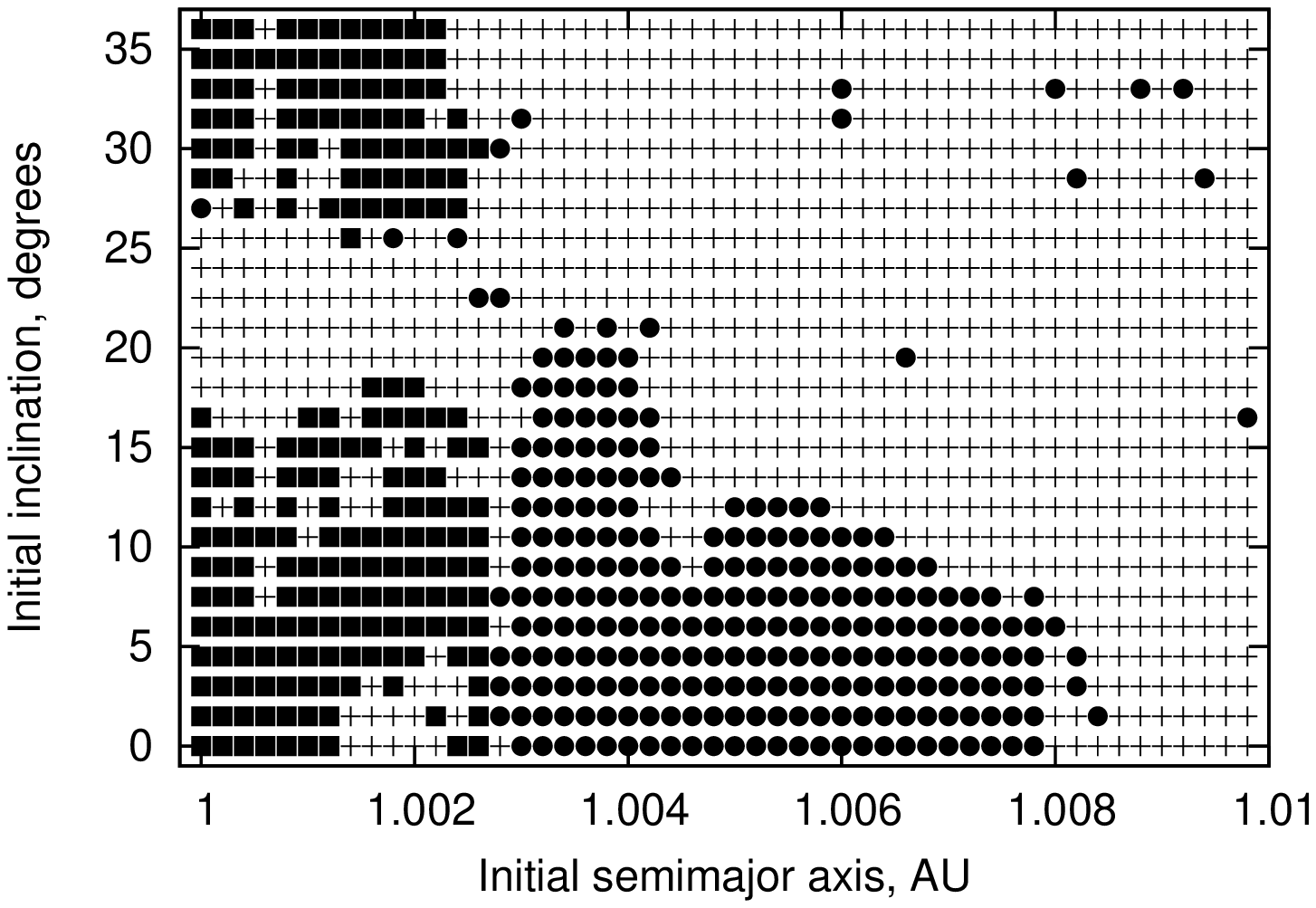}
\label{10Myr}
\end{minipage}
\hspace{0.1truein}
\begin{minipage}{3.3truein}
\centering
(b)\includegraphics[scale=0.5, angle=270]{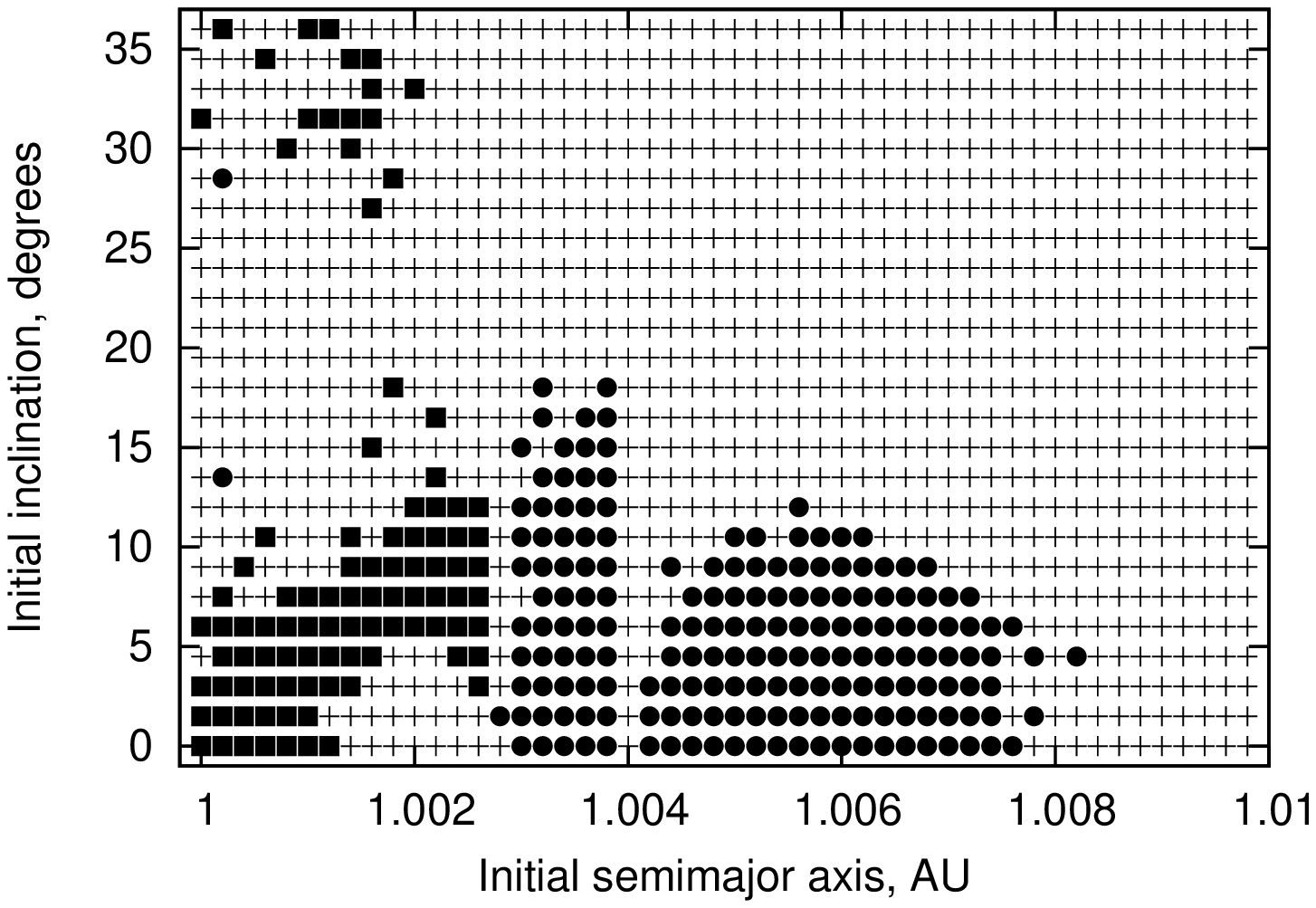}
\label{100Myr}
\end{minipage}
\begin{minipage}{2.8truein}
\centering
(c)\includegraphics[scale=0.5, angle=270]{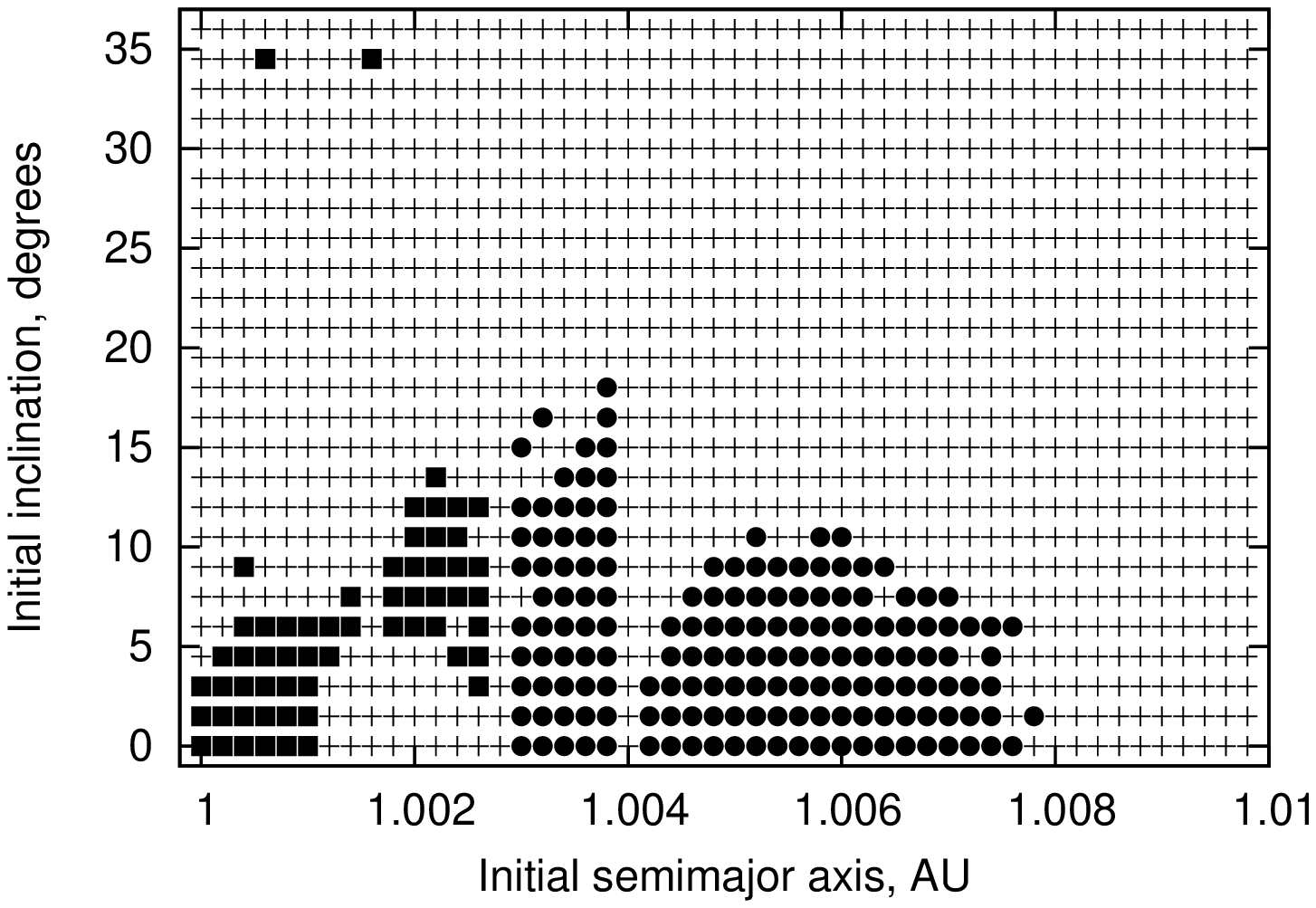}
\label{300Myr}
\end{minipage}
\hspace{0.1truein}
\begin{minipage}{3.3truein}
\centering
(d)\includegraphics[scale=0.5, angle=270]{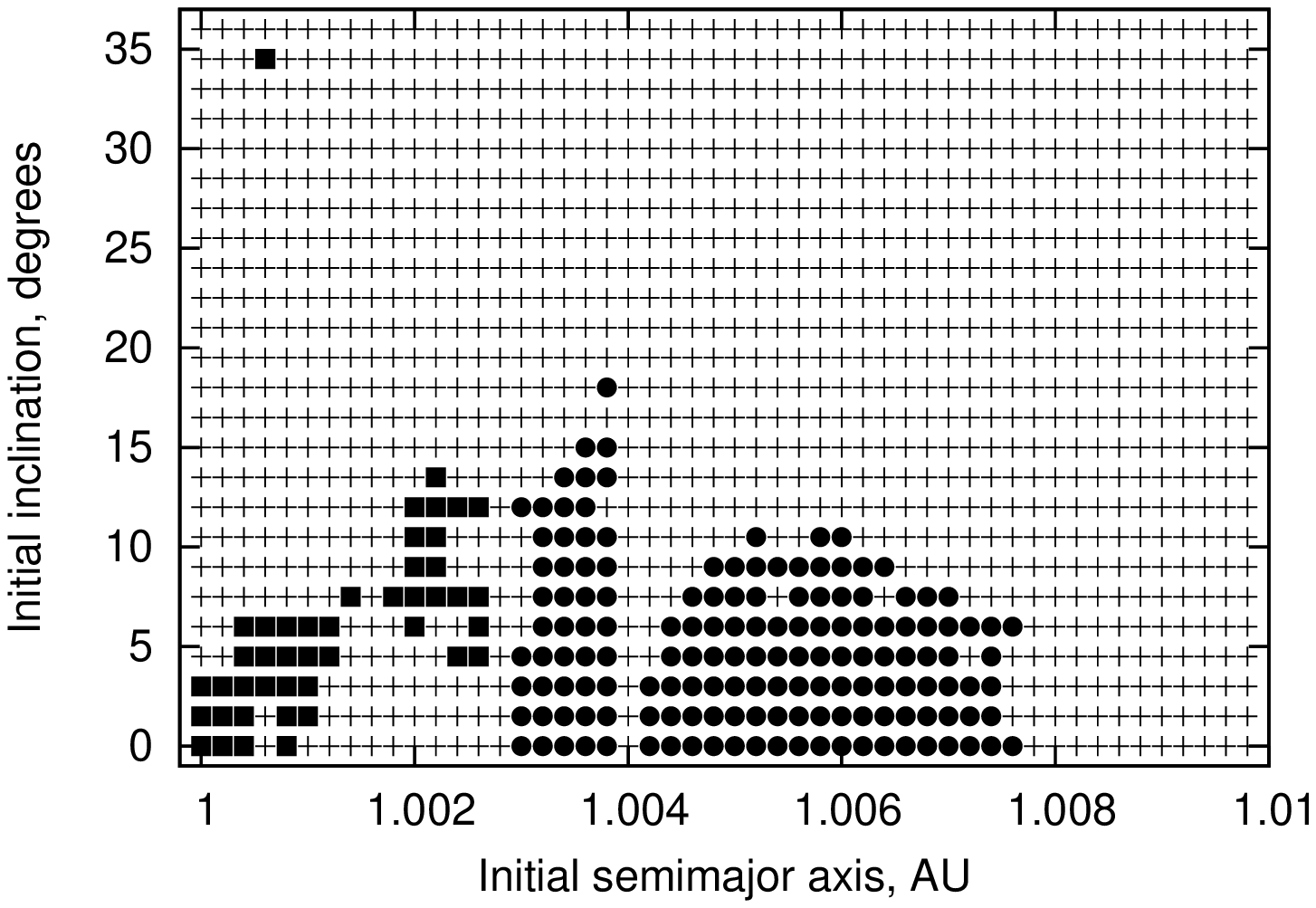}
\label{700Myr}
\end{minipage}
\centering
\caption{Remaining stable Trojans (squares) and horseshoe coorbitals (circles) of Earth after (a) 10 Myr, (b) 100 Myr, (c) 300 Myr and (d) 700 Myr. Pluses indicate unstable particles. At t=0, all orbits had e=0.05 and the same longitude of the node and pericenter as Earth.}   
\label{earth}
\end{figure*}

The initial motivation for this study was the question of the source of late lunar impactors, usually termed the Late Heavy Bombardment \citep{ccg07}. The formation of the Moon's giant Imbrium basin 3.85 Gyr ago, accompanied by many other smaller impacts, indicates that a population of impactors was present in near-Earth space about 700 Myr after the formation of the planets. While non-coorbital inner Solar System planetesimals should have been mostly cleared out by then \citep{bot07}, Eq. \ref{tau} indicates that horseshoe coorbitals may linger for a Gyr or so. Eq. \ref{tau} gives the horseshoe lifetimes of 1.6 Gyr for Earth and 0.7 Gyr for Venus, the right order of magnitude to explain the Late Heavy Bombardment \citep[and consistent with previous numerical results;][]{tab00}. Furthermore, \citet{sch05} found that Venus Trojans\footnote{Here we use "Trojan" to mean a tadpole coorbital of any celestial body.} may be also unstable on Gyr timescales. While the question of late impactors on the Moon has motivated our original numerical experiments (Section 2), our study was eventually broadened to address the intrinsic stability of horseshoe coorbitals in general (Section 3).

\section{Coorbitals of Earth and Venus}


\begin{figure*}
\begin{minipage}{2.8truein}
\centering
(a)\includegraphics[scale=.5, angle=270]{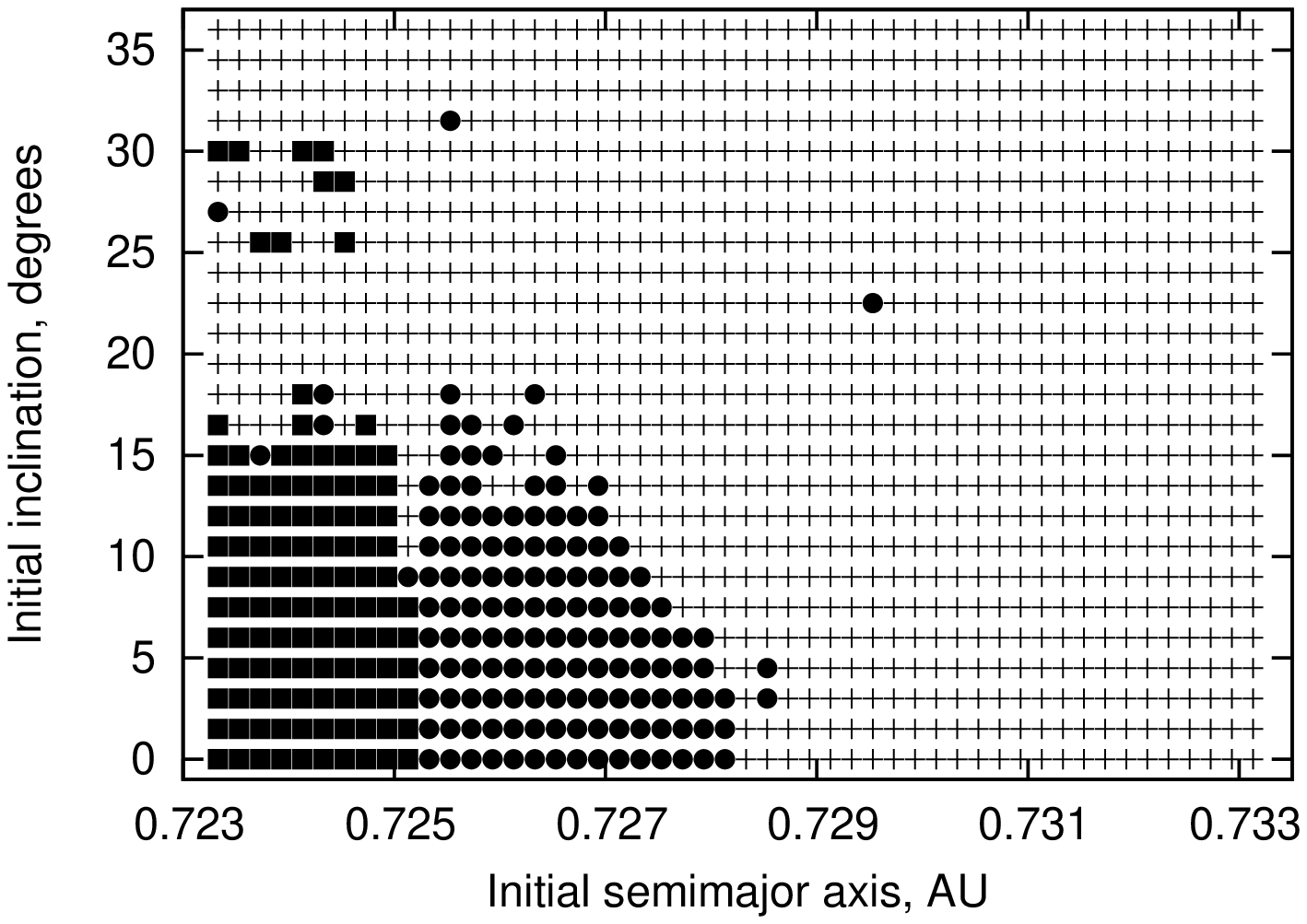}
\label{v10Myr}
\end{minipage}
\hspace{0.1truein}
\begin{minipage}{3.3truein}
\centering
(b)\includegraphics[scale=0.5, angle=270]{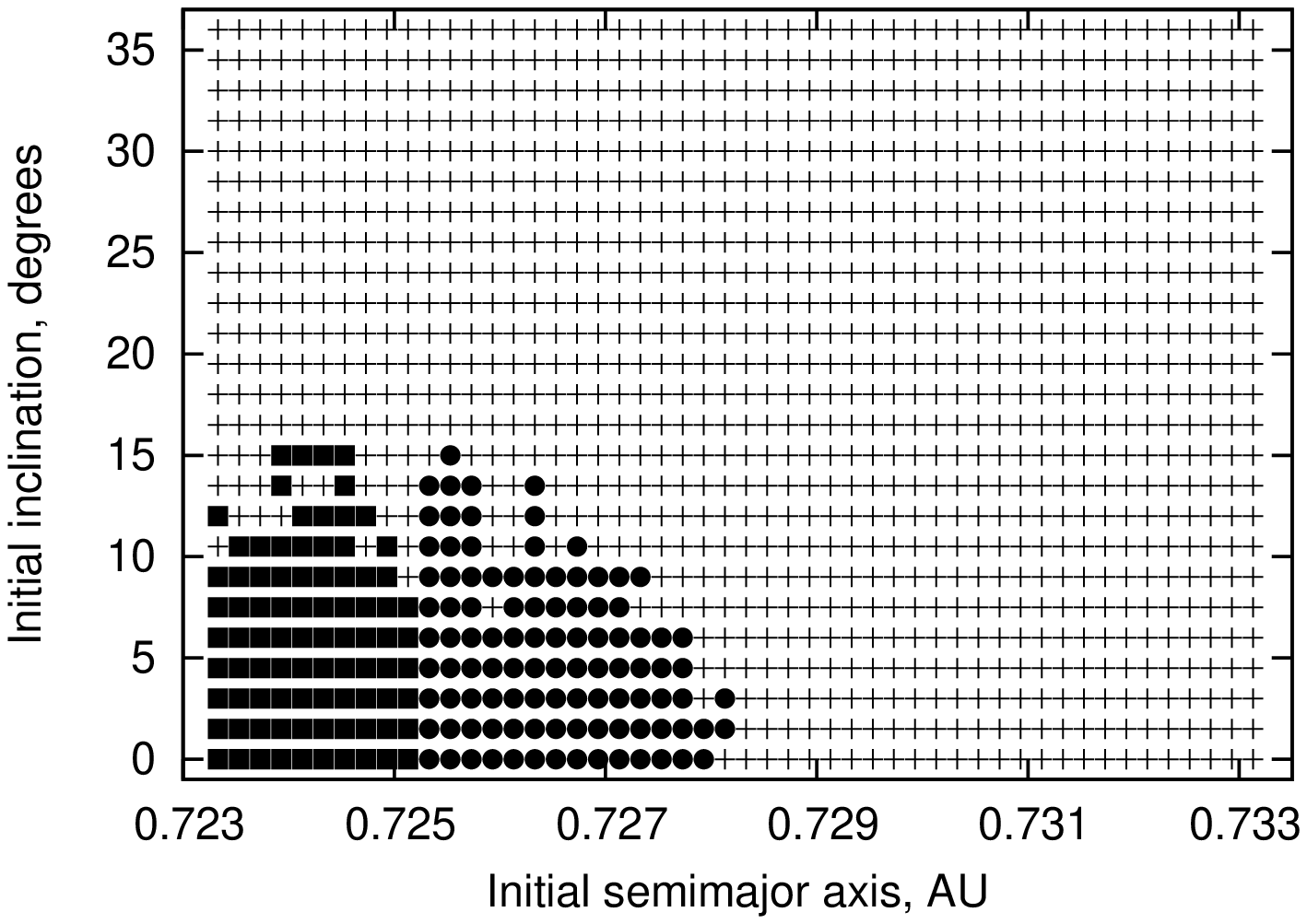}
\label{v100Myr}
\end{minipage}
\begin{minipage}{2.8truein}
\centering
(c)\includegraphics[scale=0.5, angle=270]{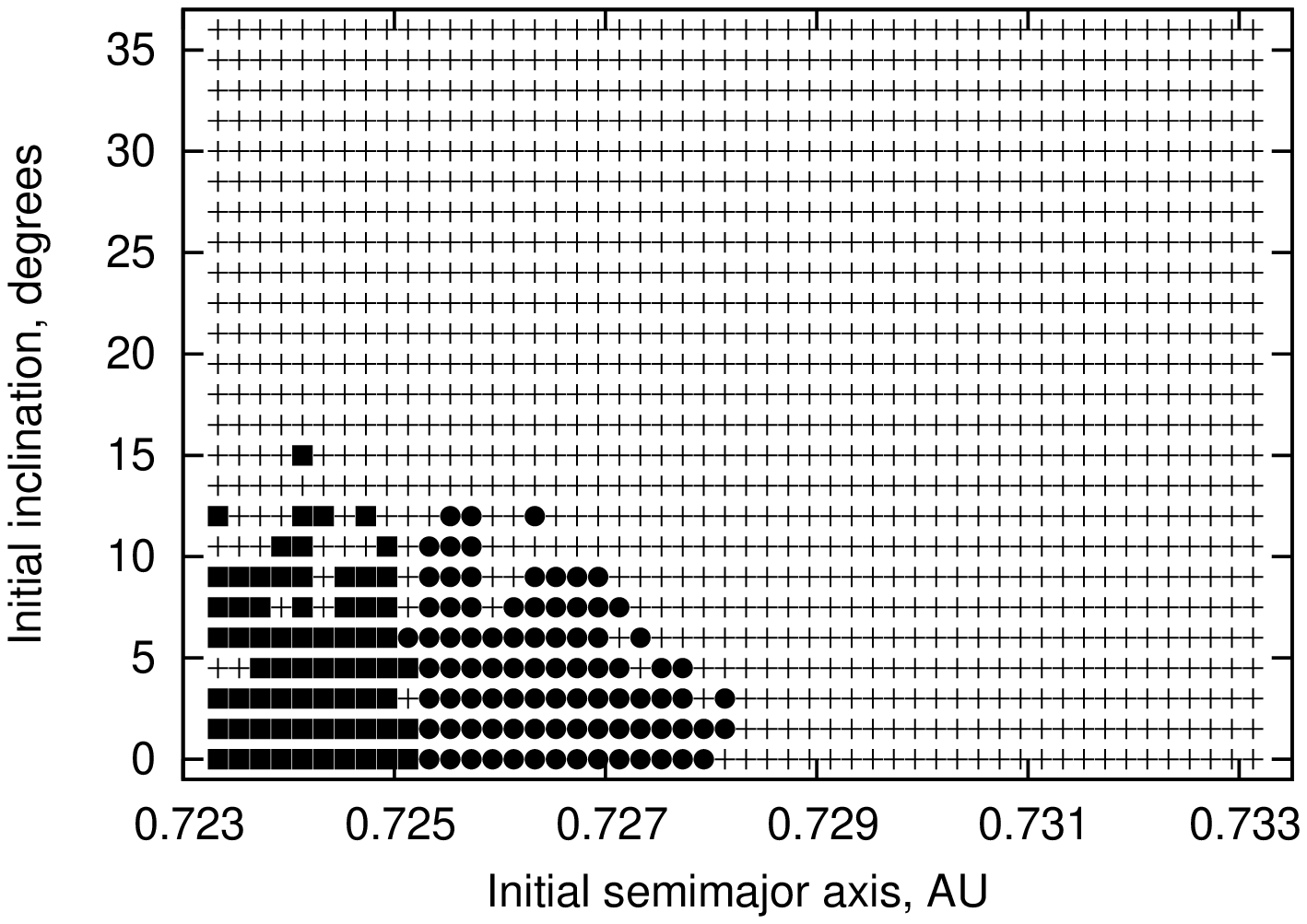}
\label{v300Myr}
\end{minipage}
\hspace{0.1truein}
\begin{minipage}{3.3truein}
\centering
(d)\includegraphics[scale=0.5, angle=270]{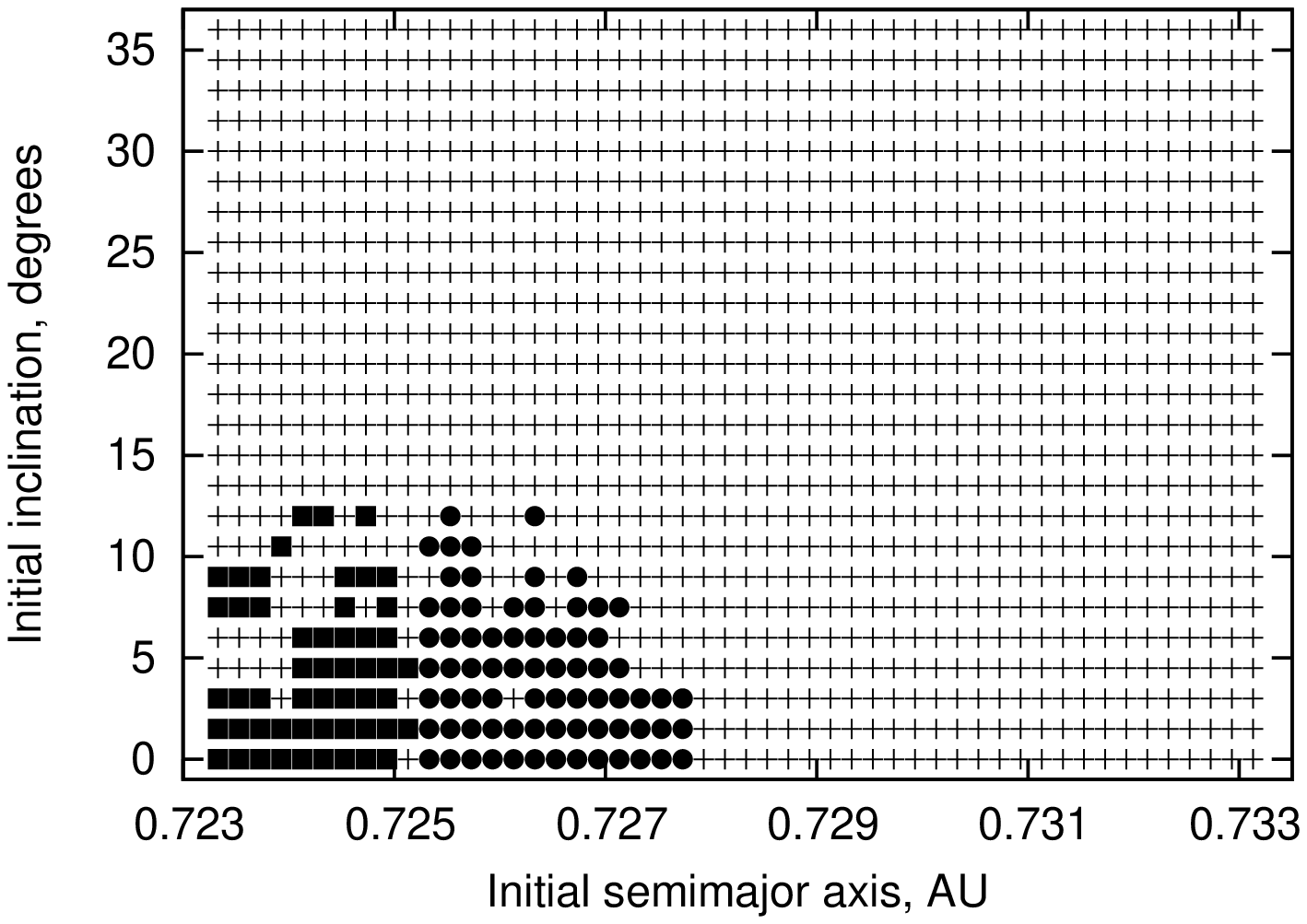}
\label{v700Myr}
\end{minipage}
\centering
\caption{Remaining stable Trojans (squares) and horseshoe coorbitals (circles) of Venus after (a) 10 Myr, (b) 100 Myr, (c) 300 Myr and (d) 700 Myr. Pluses indicate unstable particles. At t=0, all orbits had e=0.05 and the same longitude of the node and pericenter as Venus. This integration was continued to 1Gyr with little change from panel (d).}   
\label{venus}
\end{figure*}

As we were initially interested in both horseshoe and tadpole coorbitals, we decided to place our test particles 60$^{\circ}$ ahead of the planet in longitude, consistent with both types of trajectories. One thousand particles were spaced on a uniform grid, with 25 different starting inclinations and 40 different initial semimajor axes. For both Earth and Venus coorbitals, the semimajor axis range was 0.01~AU, starting at the planet's mean distance, and the inclination range was $0-36^{\circ}$ (measured from the planet's orbital plane, with the same node). This range of initial distances covers about one Hill radius from the planet's own orbit, which is the expected size of the coorbital region (see Section 3). All particles had initial eccentricities of 0.05 and the same longitude of pericentre as their parent planet. The eight planets were started on their initial positions for January 1st, 2000, and their initial vectors and masses were obtained from JPL's HORIZONS online service. To integrate the particles' orbits, we used the Swift-RMVS4 integrator (provided by Hal Levison), based on previous successful versions of Swift \citep{lev94}. Swift-RMVS4 is a "democratic" symplectic integrator \citep[cf. ][]{wis91} that uses canonical heliocentric coordinates and is capable of resolving close encounters between test particles and planets, and the RMVS4 variant is notable for planets being insensitive to the presence of test particles (in previous versions, systems having identical planets but different massless particles would slowly diverge). With RMVS4, the whole grid would experience the same planetary perturbations, even though the computation is divided between 40 different processors. We used a time step of 3 days for all runs. The Earth and Venus grids were integrated for 700 Myr and 1 Gyr, respectively. Particles that collided with the Sun or a planet, or reached 100~AU from the Sun, were removed from the simulation. All computations in this section were completed on Harvard University's computing cluster "Odyssey". 

\begin{figure}
\includegraphics[scale=0.55, angle=270]{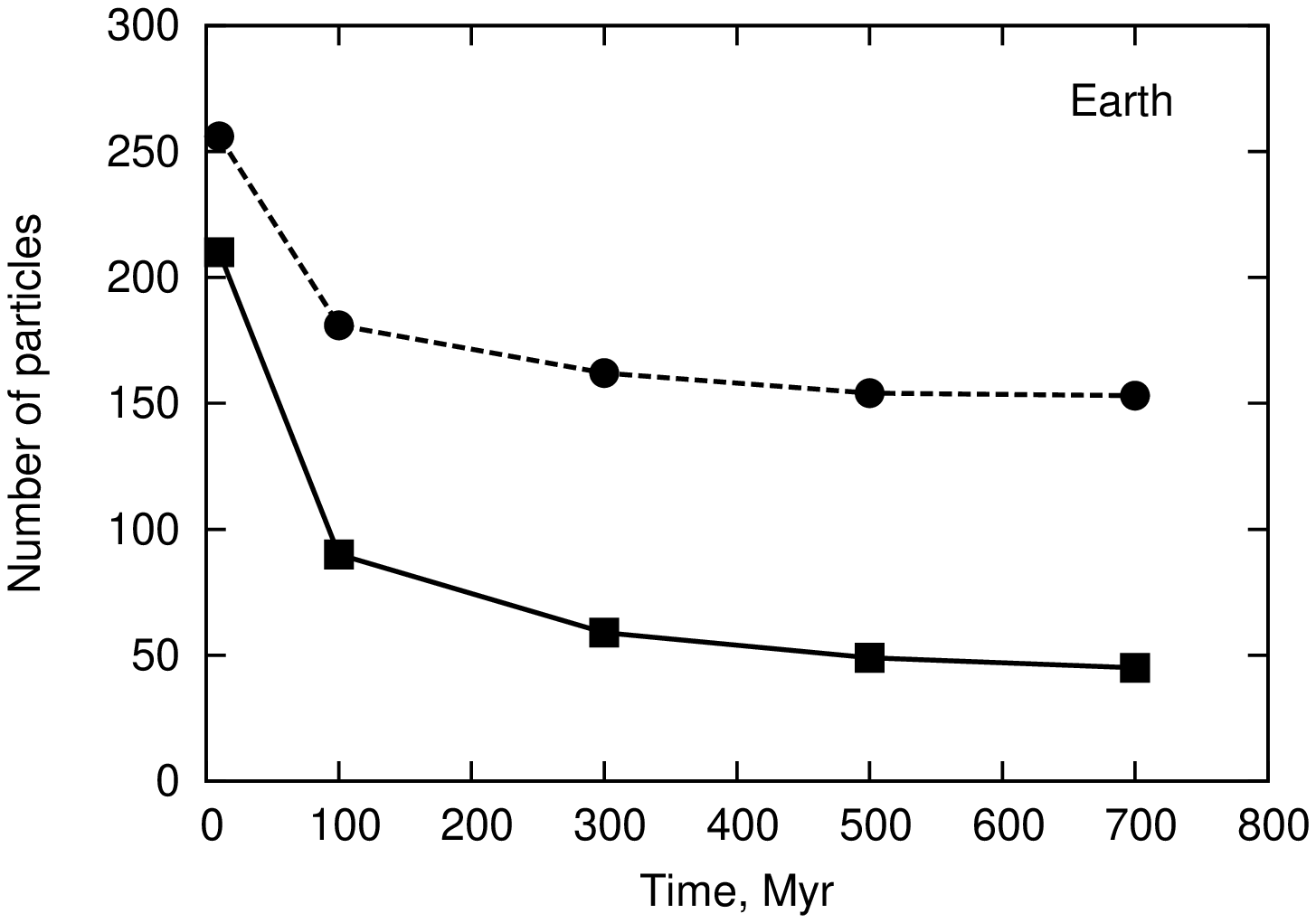}
\includegraphics[scale=0.55, angle=270]{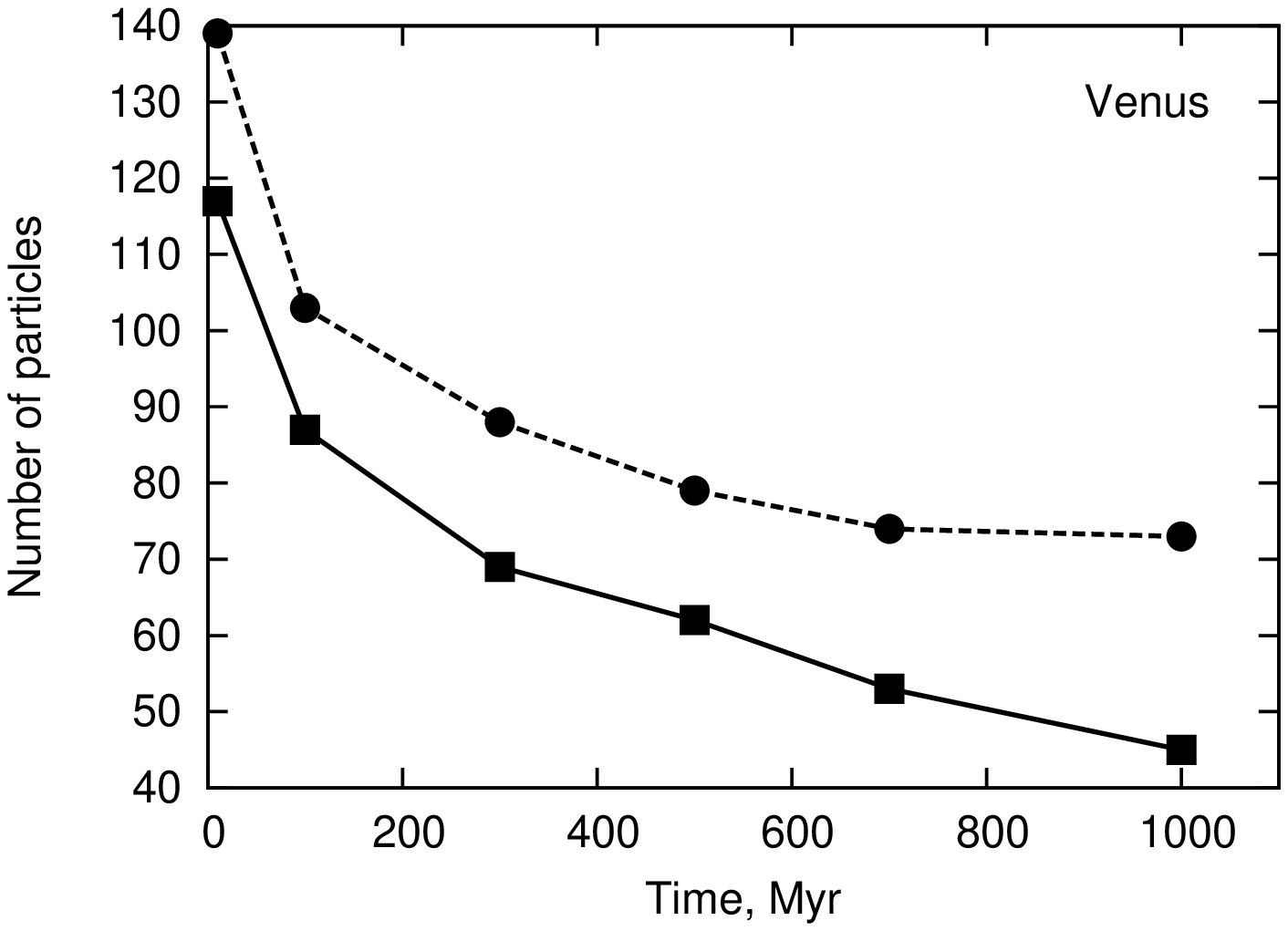}
\caption{The number of surviving particles on tadpole (squares) and horseshoe (circles) orbits as a function of time, for the coorbitals of Earth (top) and Venus (bottom).}
\label{decay}
\end{figure}

Fig. \ref{earth} shows the results of our integrations of Earth coorbitals. Panels a-d show the distribution of surviving horseshoe and tadpole objects on the grid of initial conditions, at 10, 100, 300 and 700 Myr, respectively. Fig. \ref{venus} shows the results for the Venus coorbital grid, in the same format. The inner part of both planets' coorbital region is taken up by Trojan librators, while the outer part is populated by horseshoe coorbitals, as expected \citep{md99}.\footnote{Particles found over more than $180^{\circ}$ of mean longitude in the frame rotating with the planet's mean motion were considered to be on horseshoe orbits, while those confined to a smaller range of longitudes were assumed to be Trojans.} In both cases, Trojans appear to be unstable at specific inclinations, an effect of secular resonances \citep{bra02, sch05}. As found before by \citet{sch05} for Venus, horseshoe coorbitals are most stable at low inclinations. Contrary to our expectations from Eq. \ref{tau}, many horseshoe coorbitals appear to be stable at longer timescales, and there is little change in the number of stable horseshoe objects between 300~Myr (panels c) and 700~Myr (panel d) in both figures. The evolution of the two grids is summarized in Fig. \ref{decay}, which shows the number of survivors in each population over time. The number of surviving horseshoe coorbitals of both planets seems to plateau after a couple hundred Myr, and there are barely any escaping horseshoe particles after 500 Myr. Trojans appear to have more of a long-lived but unstable tail, but also appear incapable of adding sufficiently to the Late Heavy Bombardment. 

It is interesting to compare our results to the orbit of the most long-lived temporary horseshoe coorbital of Earth, 2010~SO16 \citep{chr11}. It has a somewhat higher eccentricity that our grid (0.075 vs. 0.05), and its semimajor axis semi-amplitude (0.004~AU) and inclination ($14.5^{\circ}$) place it on the edge of the peninsula of instability near the centre of Fig. \ref{earth}a, which becomes a vertical gap in panels b-d. Therefore, our integrations are consistent with a relatively long-lived, but ultimately unstable nature of this object \citep{chr11}.

Our results diverge from those of \citet{sch05} for Venus Trojans. They identified a subset of thirty most stable low-inclination Trojans of Venus and followed them for 1 Gyr, during which time all of the particles became unstable. Figs. \ref{venus} and \ref{decay} show that there is still a significant number of surviving Venus Trojans at 1 Gyr, which were unlikely to be missed by \citet{sch05}. Apart from the timescale, we also find different profile of Trojan loss for Venus: our results show a slow exponential-like decay, while the population drops approximately linearly for \citet{sch05}. Despite the small numbers of particles, we find that our results are significantly different statistically. \citet{sch05} also used Swift, so it is unlikely that the difference is an artifact of our integration schemes.

We hypothesize that there is a hidden physical parameter that makes our Venus Trojans more stable than those of \citet{sch05}. We explored the possibility that the divergence of planetary orbits due to chaos \citep{las94, las08} may lead to different incarnations of the Solar System in our simulations, which could have consequences for stability of coorbitals. In particular, the strength of the normal mode associated with the eccentricity of Mercury can vary significantly over time, possibly also affecting the secular behavior of Venus Trojans. In order to test this, we integrated a subset of relatively stable coorbitals of both Earth and Venus ten times with slightly different initial conditions (achieved by changing the distance of Earth from the Sun by a very small amount). While the integrations did diverge, there was no real difference in the stability of coorbitals over 500 Myr. We therefore think that the secular state of the system is not the reason for the discrepancy between the results of \citet{sch05} and our own. One remaining possibility is that planetary mean-motions may be slightly different between our runs and those of \citet{sch05}. We did not take care to avoid "cold start" issues and match the mean motions of planets precisely to the observed values (as symplectic integrators take large steps, correct initial conditions can generate an orbit that is slightly different from the correct one). Therefore, if some combination of planetary motions is near-resonant with Trojan librations \citep[cf. ][]{nes02}, it would be very sensitive to small changes in the planets' orbital periods, and may have very different effects in our simulations and those of \citep{sch05}. However, this issue is too tangential to explore here in more detail, as we are mainly concerned with the stability of horseshoe coorbitals. We therefore leave this discrepancy unresolved.
 
\section{Coorbitals in Single-Planet Systems}

\begin{figure}
\includegraphics[scale=0.55, angle=270]{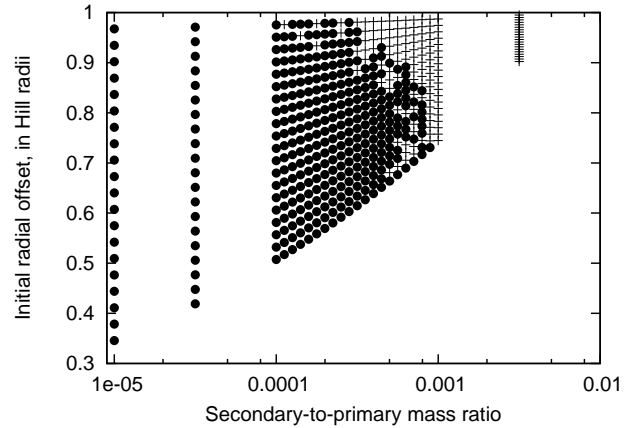}
\caption{Initial radial displacement $\Delta a$ (in planet's Hill radii) of stable (circle) and unstable (pluses) particles as a function of planet mass in our idealized one-planet, $10^8$-orbit simulations. For each planet mass, twenty particles were evenly distributed between the Trojan-horseshoe boundary and one Hill radius beyond the planet's orbital distance, $60^{\circ}$ ahead of the planet in longitude. Most (but not all) of these orbits are horseshoes.}
\label{coorb}
\end{figure}

While the coorbitals may not be a source of late lunar impactors, we are still interested in the issue of their intrinsic stability. Some horseshoe coorbitals of Earth and Venus appear to be long-term stable, in defiance of Eq. \ref{tau}. In order to study their intrinsic stability, we decided to run a number of numerical experiments in planar systems with only one planet on a circular orbit. By eliminating other planets we can make sure that any instabilities are not due to their perturbations, and also significantly speed up our integrations. First we ran seven simulations in which our lone planet had mass of $\mu=10^{-2+k\times0.5}$, where $k=1-6$. Twenty test particles were placed $60^{\circ}$ ahead of the planet, uniformly spaced between the Trojan-horseshoe boundary at $\Delta a_1= a_0 \sqrt{8/3 \mu}$ and the Hill radius $\Delta a_2= a_0 (\mu/3)^{1/3}$ (the coorbital region is actually about 20\% wider than a Hill radius, but here we are only interested in the most stable particles). The planet was assumed to orbit a solar-mass star at 1~AU, and the timestep was 10~days. All particles were followed for $10^8$ orbits. Results are plotted in Fig. \ref{coorb}. 

The boundary between stable and unstable particles is not clear-cut in either initial semimajor axis or planet mass. As the mass range $\mu=10^{-4} - 10^{-3}$ was found to contain the stability-instability transition, we filled in that region with more closely spaced integrations. Note that some of the innermost stable particles along the horseshoe-Trojan boundary actually follow tadpole orbits. In any case, it appears that all horseshoe coorbitals become unstable once the planet mass reached about one Jupiter mass ($\mu\simeq0.001$). After we obtained this result, we realized that \citet{lau02} reached a similar conclusion when they studied horseshoe configurations of two equal-mass planets. They found that two Saturn-sized planets can make a stable pair (at $\mu=3 \times 10^{-4}$ each), but more massive ones cannot. It appears that the stability boundary depends only on the total mass in two bodies, regardless of how it is distributed. This is, in fact, expected from the equations of motion of the system which predict dependence on only the sum of the coorbital masses \citep{yod83}. However, given how "fuzzy" the boundary appears in Fig. \ref{coorb}, it is very likely that some other parameter of the horseshoe contributes to its stability.

\begin{figure}
\includegraphics[scale=0.55, angle=270]{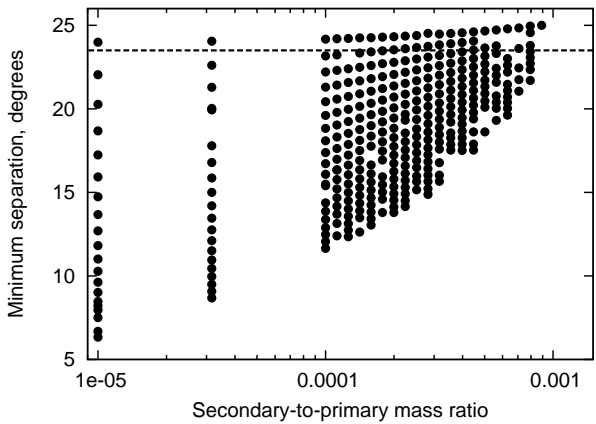}
\caption{Smallest distance in mean longitude from the planet for stable particles with initial conditions plotted in Fig 4. The particles were integrated for $10^8$ orbits. The theoretical boundary between tadpole and horseshoe orbits at closest approach (23.5$^{\circ}$) is plotted by the dashed line. Most (but not all) particles above the line are on tadpole orbits.}
\label{degrees}
\end{figure}

In Fig. \ref{degrees} we plot the minimum approach angle of stable particles' guiding centres against the planet's mass.\footnote{The guiding centre is a point that has the same semimajor axis and mean longitude as the particle itself, but with zero orbital eccentricity. In a rotating reference frame, the guiding centre is the centre of the particle's epicycle.} It is now clear that stable horseshoe particles' guiding centres appear to be restricted to a relatively well-defined area which disappears for $\mu \gtrsim 10^{-3}$. The upper boundary of this region is almost independent of planet mass, and corresponds to the Trojan-horseshoe dividing line. \citet{md99} show how, when crossing the planet's heliocentric distance at the closest approach to the planet, the guiding centres of horseshoe particles must always pass closer than 23.5$^\circ$ of the planet, while the Trojans stay further away. Our integrations agree with this result, as most stable particles above the dashed line in Fig. \ref{degrees} are actually on tadpole orbits (they were initially placed exactly on the Trojan-horseshoe boundary). 

\begin{figure}
\includegraphics[scale=0.55, angle=270]{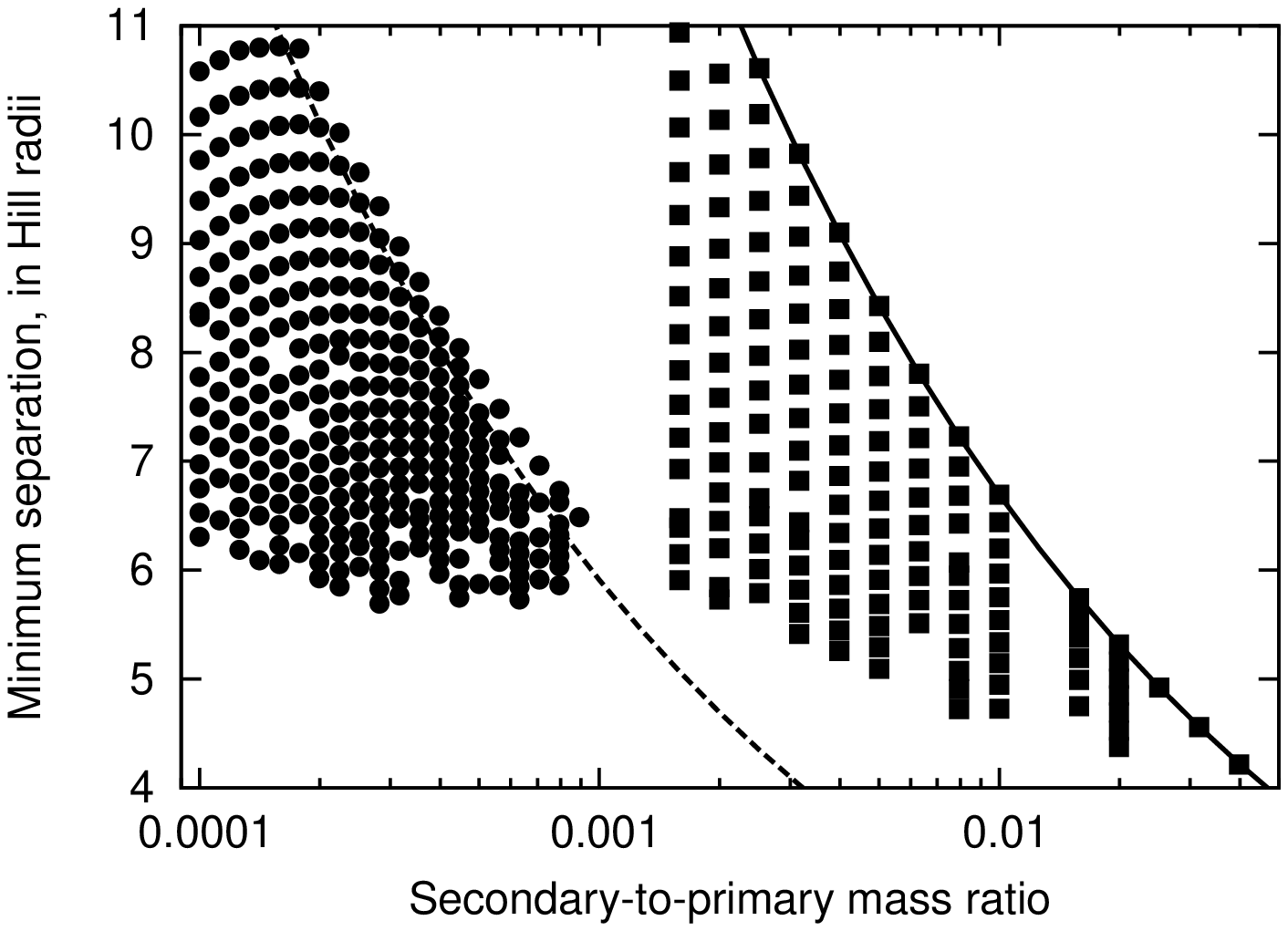}
\includegraphics[scale=0.55, angle=270]{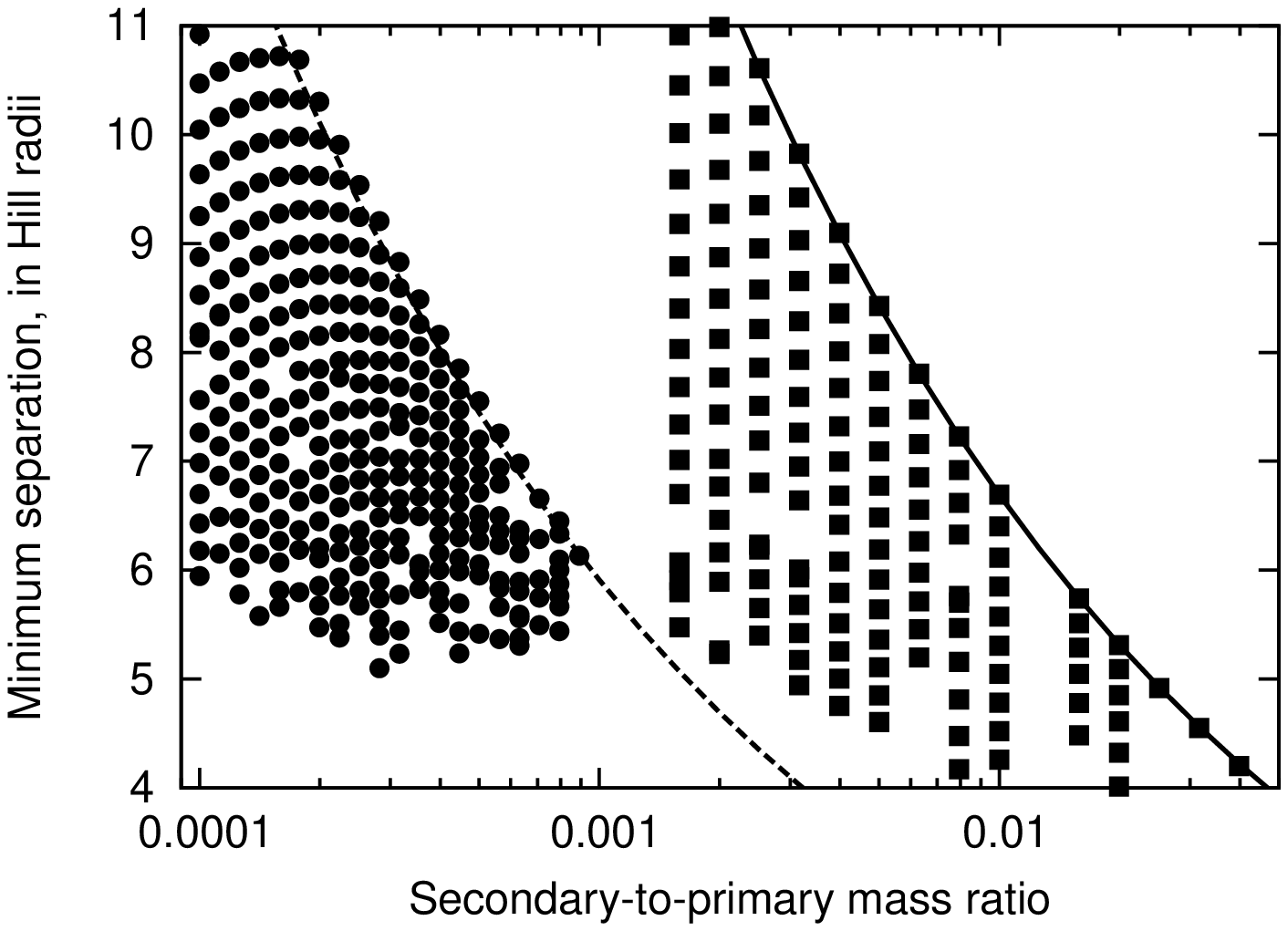}
\caption{Smallest distance from the planet (in planetary Hill radii) for the stable particles' guiding centres (top panel) and the particles themselves (bottom panel). Circles plot stable particles from Figs. 4 and 5 and squares plot a separate set of stable Trojans of higher-mass planets. Dashed and solid lines plot the distance of the tadpole-horseshoe boundary and the triangular Lagrange point, 23.5$^{\circ}$ and 60$^{\circ}$ from the planet in longitude, respectively. No stable horseshoe coorbitals approach within 5 Hill radii of the planet, while Trojans of massive planets can approach within 4 Hill radii.}
\label{hill}
\end{figure}

The bottom boundary of the stable region is curved, and appears to be a relatively weak function of planet mass. As planetary encounters usually scale with the Hill radius of the planet, we re-plotted the same data in Fig. \ref{hill}, where the minimum distance between the particle and the planet was expressed in Hill radii ($R_H=a_0 (\mu/3)^{1/3}$), rather than degrees. The minimum distance for stable horseshoe coorbitals appears to have a lower boundary at 5~$R_H$, which keeps their guiding centres at least 5.5~$R_H$ from the planet. This seems to be the determining factor for the lack of stability for horseshoes of Jupiter-mass and larger planets: even the innermost horseshoe trajectory for planets of this size experiences a relatively close approach to the planet at the horns of the horseshoe. This critical mass for horseshoe stability ($\mu \simeq 1/1200$) is the real explanation for the instability of Jupiter horseshoe coorbitals \citep{sta08}, and the stability of low-mass systems like Janus and Epimetheus, rather than Eq. \ref{tau}. Furthermore, as noted by \citep{der81a} and demonstrated in Fig. \ref{coorb}, a larger part of the coorbital region is taken by horseshoes for smaller secondary mass.

This 5~$R_H$ (or 5.5~$R_H$ for guiding centres) minimum close-approach distance applies only to horseshoe coorbitals. We also integrated a number of Trojan orbits for planets of larger mass (squares Fig. \ref{hill}) and found that Trojans of high-mass planets can approach within 4 $R_H$ of the planet and still be stable over $10^8$ orbits. It is not clear if the Trojans are intrinsically more stable than horseshoes when subjected to the same planetary perturbations, or the minimum distance simply decreases with increasing planet mass. Not coincidentally, the stable region for Trojans in Fig. \ref{hill} disappears at about the same mass ratio at which the Lagrange point becomes unstable, $\mu=0.04$ \citep{md99}.

\begin{figure}
\includegraphics[scale=0.65, angle=0]{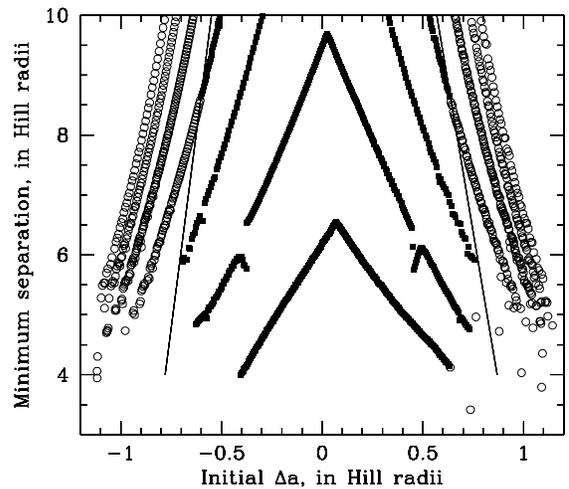}
\caption{Closest approach to planet in Hill radii against the initial radial displacement (from a Lagrange point) for coorbital particles stable over $10^8$~orbits. Each linear feature is comprised of stable horseshoe (open circles toward the sides of the plot) and tadpole (small squares near the centre of the plot) coorbitals for one particular secondary mass.  The innermost triangular peak corresponds to $\mu=$0.01 and the outermost to $\mu=10^{-6}$, and the others are for masses spaced $\Delta \log(\mu)=0.5$ apart. The thin solid line plots the $23.5^\circ$ horseshoe-Trojan boundary. These initial conditions were integrated using HNBody symplectic integrator \citep{rau02, rau12}, and the minimum distance from the planet was calculated at every timestep. }
\label{doug}
\end{figure}

In order to check that our results are not integrator-dependent, we ran a second set of simulations with some of the same initial condition propagated using the HNBody symplectic integrator \citep{rau02, rau12}. Fig. \ref{doug} shows the results of these integrations, which also ran for $10^8$~orbits, but the closest approach distance was calculated every timestep (in Fig. \ref{hill} the minimum distance was only computed for output times $0.1$~Myr apart). While some stable horseshoe coorbitals do occasionally approach as close as 4~$R_H$ to the planet, the region of contiguous horseshoe stability ends at 5~$R_H$, in agreement with Fig. \ref{hill}.  

\section{Conclusions}

Using direct numerical integrations, we find that a substantial number of horseshoe coorbitals of Earth and Venus appear to be long-term stable. Some of the low-inclination Trojans of these two planets are also stable over 1 Gyr, but the escapes from the stable region continue throughout the simulations, likely due to secular interactions with the planets \citep{bra02, sch05}. More work is needed to ascertain if any of the Trojans of Earth and Venus are stable over the Solar System's lifetime, and if not, what are the timescales and exact mechanisms of their instability. In any case, the lack of observed primordial coorbitals of Earth or Venus, together with our finding of their relative stability, indicate that these objects were not an important contribution to late bombardment on Earth and the Moon.

We also integrated a number of horseshoe coorbitals in planar, circular one-planet systems. We find that only planets smaller than about one Jupiter mass ($\mu=1/1200$ or smaller) can have stable horseshoe coorbitals. The main reason for this stability boundary is that the horns of the horseshoe approach too close to the planet (in terms of the planet's Hill radii) for more massive planets. This result is in agreement with the work of \citet{lau02} on equal-mass planet pairs, the known instability of Jupiter horseshoes, and the long-term stability of the only observed horseshoe pair, Janus and Epimetheus.

\section*{Acknowledgments}

M. \'C. thanks Hal Levison for providing his Swift-RMVS4 numerical integrator. This work was done while M. \'C. was Daly Postdoctoral Fellow at Harvard University and Clay Postdoctoral Fellow at the Smithsonian Astrophysical Observatory. Numerical integrations were completed on Harvard University's Odyssey cluster and at the University of Maryland. We thank Carl Murray for a helpful review.

\bsp

\label{lastpage}

\end{document}